\documentstyle[12pt,epsf]{article}
\newcommand{\be}{\begin{equation}}
\newcommand{\ee}{\end{equation}}
\newcommand{\bea}{\begin{eqnarray}}
\newcommand{\eea}{\end{eqnarray}}
\newcommand{\nn}{\nonumber}

\newcommand{\NP}{Nucl. Phys.}

\newcommand{\PRL}{Phys. Rev. Lett.}
\newcommand{\PL}{Phys. Lett.}
\newcommand{\PR}{Phys. Rev.}
\def\art{\@ifnextchar[{\eart}{\oart}}
\def\eart[#1]#2#3#4#5#6{{\rm #2}, {\em #3 \rm #4} {\rm (#6) #5 ({\em #1})}}
\def\hepart[#1]#2{{\rm #2, \em#1}}
\newcommand{\oart}[5]{{\rm #1}, {\em #2 \rm #3} {\rm (#5) #4}}

\newcommand{\vn}{{\vec{n}}}

\newcommand{\hg}{{\hat{g}}}

\newcommand{\tA}{{\widetilde{A}}}

\newcommand{\tF}{{\widetilde{F}}}
\newcommand{\th}{{\widetilde{h}}}
\newcommand{\tp}{{\widetilde\phi}}

\begin{document}

\def\htil{\widetilde h}
\def\phitil{\widetilde \phi}
\def\del{\delta}
\def\mt{m_t}
\def\mpzero{M_{\pzero}}
\def\mev{~{\rm MeV}}
\def\gev{~{\rm GeV}}
\def\gam{\gamma}
\def\lsim{\mathrel{\raise.3ex\hbox{$<$\kern-.75em\lower1ex\hbox{$\sim$}}}}
\def\gsim{\mathrel{\raise.3ex\hbox{$>$\kern-.75em\lower1ex\hbox{$\sim$}}}}
\def\tr{{\rm {tr}}}
\def\LL{{\cal L}}
\def\f{\frac}
\def\vfi{\varphi}
\def\gt{\tilde g}
\def\ct{{\tilde c}_\theta }
\def\st{{\tilde s}_\theta }
\def\sb{s_\beta }
\def\cb{ c_\beta }
\def\eps{\epsilon}
\def\s{s_\theta }
\def\c{c_\theta }
\def\sz{\sum_{n=0}^\infty}
\def\su{\sum_{n=1}^\infty}
\def\de{\partial}
\def\half{{1\over 2}}
\def\ldel{L}
\def\bW{{\bf W}}
\def\bY{{\bf Y}}
\def\k{\kappa}
\def\DD{{\cal D}}
\font\fortssbx=cmssbx10 scaled \magstep2

$\vcenter{
\hbox{\fortssbx University of Florence}
}$
\hfill
$\vcenter{
\hbox{\bf DFF-362/9/00}
}$
\medskip
\begin{center}
{\Large\bf\boldmath Non Linear Breaking of the Electroweak Symmetry and
Large Extra Dimensions}
\rm
\vskip1pc
{\Large  D. Dominici$^{a,b}$}
\end{center}
\begin{center}
\vspace{5mm}
{\it{$^a$Dipartimento di Fisica, Universit\`a di Firenze, I-50125
Firenze, Italia
\\
$^b$I.N.F.N., Sezione di Firenze, I-50125 Firenze, Italia\\
}}
\end{center}
\bigskip
\begin{abstract}
 We consider the coupling  to gravity in $4+\delta$ dimensions
of a non linear  
electroweak symmetry  breaking sector,
 with $\delta$ compactified dimensions, and 
derive an effective lagrangian by integrating
over the KK excitations of the graviton and of the dilaton.
The effective chiral lagrangian describes the interactions of the
electroweak gauge bosons and the goldstone bosons from
the electroweak breaking.
\noindent
\end{abstract}
\newpage

\section{Introduction}
There is been recently a growing interest in theories in $4+\delta$
dimensions with $\delta$ compactified dimensions  and an effective
Planck scale  $M_{Pl(4+\delta)}$ which can be lower than the Planck scale
\cite{wit,ark}.
In these models gravity propagates in $4+\delta$ dimensions  while strong
and electroweak interactions are bounded on a three
dimensional brane.
Experimental tests of gravity exclude an effective Planck scale $M_{Pl(5)}$
down to one TeV for $\delta=1$ and a recent experiment
has also put a limit $M_{Pl(6)}\geq 3.5~TeV$ for the case $\delta=2$
\cite{wash}.
The phenomenology of these scenarios has been
largely studied \cite{gian,pesk,han}, concentrating mainly on
the KK excitations of the gravitons \cite{gian,pesk,han}
but also on KK excitations of the dilatons \cite{han,jack,gian2}.
If the gravity is brought down to one TeV scale then there could be
some intertwining between gravity and electroweak symmetry
breaking. 
A new mechanism of spontaneous symmetry
breaking induced by the Kaluza Klein (KK) excitations have been 
recently proposed \cite{jack}. The phenomenological consequences
of the graviscalar Higgs boson mixing have been
analysed \cite{jack,gian2}.
In this note we consider a non linear breaking of the
electroweak symmetry and derive an effective lagrangian by integrating
over the KK excitations of the graviton and of the dilaton.
The effective chiral lagrangian describes the interactions of the
electroweak gauge bosons and the goldstone bosons from
the electroweak breaking.

The starting point of the model is the  Einstein lagrangian
in $4+\delta$ dimensions  \cite{han,gian}
\be
\f 1 {{\hat \kappa}^2}\sqrt{- \hat g} \hat R
\ee
One  first expands around  the flat metric
\be
\hat g_{\hat\mu\hat \nu}=\eta_{\hat\mu\hat \nu}+
\hat \kappa \hat h_{\hat\mu\hat \nu}+{\cal O}(\hat \kappa^2)
\ee
($\hat \mu, \hat \nu =0,1,2,3,\cdots, 3+\delta$)
and then one performs a dimensional reduction using the following ansatz for
the metric
\be
\hat h_{\hat\mu\hat\nu}=V_\del^{-1/2}\left(
\begin{array}{cc} h_{\mu\nu}+\eta_{\mu\nu}\phi & A_{\mu i} \\
A_{\nu j} & 2\phi_{ij}\end{array}\right)
\ee
with
$\hat\kappa V_\del^{-1/2}=\kappa\equiv
\sqrt{16\pi}/M_{Pl}$, where $M_{Pl}$
is the usual Planck mass for three spatial dimensions
and $V_\del=L^\delta$ ($\hat\kappa\sim M_{Pl(4+\delta)}^{-(\delta/2 +1)}$)
and  
$\phi =\phi_{ii}$ denotes the dilaton mode.
After compactification on a $\delta$ dimensional torus $T^\delta$
one expands the fields in Fourier modes
$h^{\vec n}_{\mu\nu}$, $A^{\vec n}_{\mu i}$ and $\phi^{\vec n}_{ij}$.

The lagrangian for  
the massive $\vec n$-KK states  at the leading
order in $\hat \kappa$ is given by \cite{han,gian}

\begin{eqnarray}
{\cal L}^{\vec{n}}&=& 
{1\over2}\biggl(
\partial^\mu\th^{\nu\rho,\vn} \partial_\mu\th_{\nu\rho}^{-\vn} 
-\partial^\mu\th^\vn \partial_\mu\th^{-\vn}
-2\th^{\mu,\vn}\th_{\mu}^{-\vn} 
+ 2\th^{\mu,\vn}\partial_\mu\th^{-\vn}\nonumber\\
&&\qquad -m^2_\vn\th^{\mu\nu,\vn}\th_{\mu\nu}^{-\vn}
+m^2_\vn\th^{\vn}\th^{-\vn}\Biggr)
+\sum_{i=1}^n
(-{1\over2}\tF_i^{\mu\nu,\vn}\tF_{\mu\nu i}^{-\vn}
+ m^2_\vn\tA_i^{\mu,\vn}\tA_{\mu i}^{-\vn})\nonumber\\
&&
+\sum_{(ij)=1}^{\delta}
(\partial^\mu\tp^\vn_{ij}\partial_\mu\tp^{-\vn}_{ij}
-m^2_\vn\tp^\vn_{ij}\tp^{-\vn}_{ij})
\label{lqua}
\end{eqnarray}
where $\th_\rho^\vn= \de^\nu\th_{\nu\rho}^\vn$ and
\be
m_n^2={4\pi^2 n^2\over \ldel ^2}
\ee
 with $n^2={\vec n}^2$, $\vec n=(n_1,n_2,\ldots,n_\del)$.
For the definition of the fields 
$\th^{\vec n}_{\mu\nu}$, $\tA^{\vec n}_{\mu i}$ and $\tp^{\vec n}_{ij}$
 in terms of
$h^{\vec n}_{\mu\nu}$, $A^{\vec n}_{\mu i}$ and $\phi^{\vec n}_{ij}$
see 
\cite{han,gian}.

The fields are subjected to the constraints
\be
\de^\mu \tA^{\vec n}_{\mu i}=0,~~n_i\tA^{\vec n}_{\mu i}=0~~,
~~n_i\tp^{\vec n}_{ij}=0
\ee
The total lagrangian $\LL$ is obtained by summing $\LL^{\vec n}$
only over $\vec n$ values such that the first non-zero $n_k$
is positive. We will denote this  sum by $\sum_{\vec n>0}$ \cite{jack}.

Matter fields leave in four dimensions and their action is given by
\be
\int d^4x \sqrt{-\hat g_{ind}} {\cal L}(\hat g_{ind},S,V,F)
\label{lcoup}
\ee
for scalar (S), vector (V) and fermions (F).
Expanding the metric $(\hat g_{ind})_{\mu\nu}=\eta_{\mu\nu}+\kappa (
h_{\mu\nu}+\eta_{\mu\nu} \phi)$, the order $\kappa$ of eq.(\ref{lcoup})
is given by
\be
- \f \kappa 2 \int d^4x (h^{\mu\nu}T_{\mu\nu}+\phi T^\mu_\mu)
\ee
where
\be
T_{\mu\nu}=-\eta_{\mu\nu} {\cal L} + 2\left ( \f {\de{\cal L}}
{\de \hg^{\mu\nu}}\right )_{\hg^{\mu\nu}=\eta^{\mu\nu}}
\label{emtens}
\ee 
is the energy-momentum tensor.

The interactions of the massive $\vec n$-KK states
to the matter fields are  given by
\bea
{\cal L}_{\rm mix}^{\vec n}
&=&-{\kappa\over2}\left[\left(\htil^{\mu\nu,\vec n}+
\htil^{\mu\nu,-\vec n}\right)T_{\mu\nu}
+\omega_\del\left(\phitil^{\vec n}+\phitil^{-\vec n}\right)T_\mu^\mu\right]
\label{lmix}
\eea
where 
\be
\omega_\del=\left [2\over 3(\del+2)\right ]^{1/2}
\label{odeldef}
\ee

In conclusion we will consider the lagrangian
\be
{\cal L}=\sum_{\vec n>0}
\left [{ \cal L}^{{\vec n}}+{\cal L}_{\rm mix}^{\vec n}\right ]
\ee

Integrating
out  $\htil^{\vec n}$, $\tA^{\vec n}$ and $\phitil^{\vec n}$ fields, 
one gets an effective lagrangian  given by \cite{han,gian,jack}
\be
 {\cal L}_{\rm eff}={\kappa^2\over 4} \sum_{{\rm all}\,\vec n}\DD
[T^{\mu\nu} P_{\mu\nu;\rho\sigma} T^{\rho\sigma}
+{\omega_\del^2(\del-1)\over
2} T_\mu^\mu T_\nu^\nu ]
\ee
where
\be
\DD=\f 1 {\Box +m_n^2}
\ee
and $P_{\mu\nu;\rho\sigma}$ in the momentum space is
\bea
P_{\mu\nu;\rho\sigma}&=&\f  1 2 (g_{\mu\rho}g_{\nu\sigma} +
g_{\mu\sigma}g_{\nu\rho} -g_{\mu\nu}g_{\rho\sigma})\nn\\
&-&\f 1 {2 m_n^2}(g_{\mu\rho}k_\nu k_\sigma+
g_{\mu\sigma}k_\nu k_\rho+ (\mu\to\nu)\nn\\
&+&\f 1 6 (g_{\mu\nu}+ \f {2} {m_n^2} k_\mu k_\nu)
(g_{\rho\sigma}+ \f 2 {m_n^2} k_\rho k_\sigma)
\eea
The integration over the fields $\tA^{\vec n}$ is trivial,
because these fields decouple.
By using in the effective action the conservation of $T_{\mu\nu}$ 
(or the classical equations of motion),  one can rewrite 
${\cal L}_{\rm eff}$ as
\be
{\cal L}_{\rm eff}'= {\kappa^2\over 4}\sum_{{\rm all}\,\vec n}\DD \left(
T^{\mu\nu}T_{\mu\nu}-{1\over 3}T_\mu^\mu T_\nu^\nu
+{\omega_\del^2(\del-1)\over
2} T_\mu^\mu T_\nu^\nu\right )
\label{effhf}
\ee
The sum
$
{\kappa^2\over 4}\sum_{{\rm all}\,\vec n}\DD
$
is ultraviolet divergent for $\delta \geq 2$. Different procedures
of regularization have been suggested \cite{gian,han}. 
For instance  using
an ultraviolet cutoff $M_S$ one gets, for $M_S>> \sqrt{s}$, \cite{han}
\be
{\kappa^2\over 4}\sum_{{\rm all}\,\vec n}\DD =
{\kappa^2\over 2}\sum_{\vec n>0} \DD
\sim \f 1 
{2 M_S^4 (\delta -2)}+O(s/M_S^2)~~~~for~\delta>2
\label{dg2}
\ee
\be
{\kappa^2\over 4}\sum_{{\rm all}\,\vec n}\DD =
{\kappa^2\over 2}\sum_{\vec n>0} \DD 
\sim \f 1 {4 M_S^4}
log\frac {M_S^2}{s}+O(s/M_S^2)~~~~for~\delta =2
\label{deq2}
\ee

\section{Coupling to scalars}

For a general complex scalar field $\Phi$, we have
the following  energy-momentum tensor
\begin{equation}
T^{\rm S}_{\mu\nu}\ =\ -\eta_{\mu\nu}D^\rho\Phi^\dagger D_\rho\Phi
        +\eta_{\mu\nu}V(\Phi^\dagger\Phi)
	+D_\mu\Phi^\dagger D_\nu\Phi
	+D_\nu\Phi^\dagger D_\mu\Phi
\end{equation}
where the gauge covariant derivative is defined as
\begin{equation}
D_\mu = \partial_\mu + i g A^a_\mu T^a
\end{equation}
with $g$ the gauge coupling,
$A^a_\mu$ the gauge fields, $T^a$ the Lie algebra
generators and $V(\Phi^\dagger \Phi)$ is the potential. 

If we are interested in the non linear breaking of the 
$SU(2)\times U(1)$ symmetry it is
usual to work with the matrix 
\be
M=\sqrt{2}\left( \begin{array}{c c}
\varphi_0&-\vfi_-^*\\
\vfi_- & \vfi_0^*\\
\end{array}\right)
=\sqrt{2}\left(\Phi,-i \sigma_2 \Phi^*\right)
\ee
if $\Phi$ denotes the SM Higgs  doublet. The description of the non linear
breaking is obtained with the condition $M^\dagger M=v^2$, $v^2=1/(\sqrt 2 G_F)$
 and therefore
it is more convenient to introduce a unitary matrix $U$, such that
$M=vU$. By requiring a custodial $SU(2)$ the
effective  scalar  field lagrangian at the lowest order is then
written as \cite{app}
\bea
\LL&=& \f {v^2} 4 g_{\sigma\rho}\tr(D^\sigma U^\dagger D^\rho U) 
-\f 1 2 g_{\mu\nu}g_{\rho\sigma} \tr (\bW^{\mu\rho} \bW^{\nu\sigma})
- \f 1 2 g_{\mu\nu}g_{\rho\sigma} \tr (\bY^{\mu\rho} \bY^{\nu\sigma})
\nn\\
&+&
{\cal O} (p^4)
\label{chir}
\eea
where $D_\mu$ denotes the $SU(2)\times U(1)$ covariant derivative
\be
D_\mu U=(\de_\mu+\f{i}{2}g { W}_\mu ^a\tau^a) U-\f{i}{2}g' U
Y_\mu\tau^3
\ee
and
\be
\bW_{\mu\nu}=\f {\tau_i} {2}
(\de_\mu W_\nu^i -\de_\nu W_\mu^i -g \eps^{ijk} W_\mu^jW_\nu^k)
\ee
\be
\bY_{\mu\nu}=\f {\tau_3} {2}
(\de_\mu Y_\nu -\de_\nu Y_\mu)
\ee
Notice that the nonlinear realization of the scalar field excludes
terms of the type $R \Phi^\dagger \Phi$ \cite{gian2},
$R$ being the scalar curvature.

The energy momentum tensor, using eqs.(\ref{emtens}) and (\ref{chir}),
 can
be expressed as
\be
T_{\mu\nu}=T^{\rm S}_{\mu\nu}+T^{\rm G}_{\mu\nu}
\label{tens}
\ee
where
\bea
T^{\rm S}_{\mu\nu} &=& -\eta_{\mu\nu}  \f {v^2} {4} 
\tr[(D_\rho U)^\dagger D^\rho U]\nn\\
&+& \f {v^2} {4}  \left\{ \tr[(D_\mu U)^\dagger D_\nu U] + 
 \tr[(D_\nu U)^\dagger D_\mu U]\right \}
\eea
is the contribution from the scalar field lagrangian
and
\bea
T^{\rm G}_{\mu\nu} &=& \f 1 2 \eta_{\mu\nu}
[\tr (\bW^{\rho\sigma} \bW_{\rho\sigma}) +
\tr (\bY^{\rho\sigma} \bY_{\rho\sigma})]\nn\\
&-& 2 
[\tr (\bW_{\mu\rho} \bW_{\nu}^{~\rho}) +
\tr (\bY_{\mu\rho} \bY_{\nu}^{~\rho})]
\eea
is the contribution from the gauge field kinetic term lagrangian.
Notice that $(T^G)_\mu^\mu=0$.

\section{Effective lagrangian}

By using the eqs.(\ref{effhf}) and
(\ref{tens}) one finds the following expression for
the effective lagrangian
\be
\LL=\LL_0+\LL_{KK}
\ee 
where $\LL_0$ is given in eq.(\ref{chir}),
\be
\LL_{KK}=\LL^S+\LL^G
\ee
with
\bea
\LL^{S}&=&
\f {\k^2 v^4}{8}\sum_{\vec n>0} \DD 
\Big [(\tr[D_\mu U^\dagger D_\nu U])^2\nn\\
&+& ( -\f {1} {3} + \omega_\del^2 \f {\del -1}{2}) 
\tr(D_\rho U^\dagger D^\rho U)^2\Big ]\nn\\
&+&\f {\k^2} {2}\sum_{\vec n>0}\DD\Big [\f {v^2}{2}
\tr(D_\rho U^\dagger D^\rho U)
[\tr (\bW^{\mu\nu} \bW_{\mu\nu}) +
\tr (\bY^{\mu\nu} \bY_{\mu\nu})]\nn\\
&-& 2 v^2 \tr[D^\mu U^\dagger D^\nu U]
[\tr (\bW_{\mu\rho} \bW_{\nu}^{~\rho}) +
\tr (\bY_{\mu\rho} \bY_{\nu}^{~\rho})]\Big ]
\eea
and
\bea
\LL^{G}&=&-\f {\k^2} {2}\sum_{\vec n>0}\DD
 [(\tr (\bW^{\mu\nu} \bW_{\mu\nu}))^2+
(\tr (\bY^{\mu\nu} \bY_{\mu\nu}))^2\nn\\
&+& 2 \tr (\bW^{\mu\nu}\bW_{\mu\nu})\tr (\bY^{\mu\nu} \bY_{\mu\nu})]\nn\\
&+& 2 {\k^2} \sum_{\vec n>0}\DD
[\tr (\bW_{\mu\rho} \bW_{\nu}^{~\rho})
\tr (\bW^{\mu}_{~\rho} \bW^{\nu\rho})\nn\\
&+&\tr (\bY_{\mu\rho} \bY_{\nu}^{~\rho})
\tr (\bY^{\mu}_{~\rho} \bY^{\nu\rho})\nn\\
&+&2 \tr (\bW_{\mu\rho} \bW_{\nu}^{~\rho})
\tr (\bY^{\mu}_{~\rho} \bY^{\nu\rho})]
\eea
where now $
\sum_{\vec n>0}\DD$ is given by the expansions in eqs. 
(\ref{dg2},\ref{deq2}).
$\LL_{KK}$ is the term coming from  integrating the KK excitations
of the graviton and of the dilaton.
$\LL^S$ contains terms of dimension  four and six, while $\LL^G$ 
contains dimension eight operators.
These contributions  from KK excitations will add to other
 terms in eq.(\ref{chir}) of order $O(p^4)$ and higher order
in principle already existing.
 For simplicity
we have neglected these terms. 

 Let us now consider    the standard parametrization for 
$SU(2)_L\times SU(2)_R$ invariant Lagrangian  up to the order $p^4$, 
\begin{eqnarray}
\LL^\prime  & = &  {v^2 \over 4}
\tr (D_\rho U^\dagger
D^\rho U) \nonumber \\
&  +& \alpha_4 [\tr(D_\mu U^\dagger
D_\nu U)]^2\nonumber \\
& +&\alpha_5
 [\tr(D_\rho U^\dagger D^\rho U)]^2
\label{a4a5}
\end{eqnarray}
Assuming $\delta>2$
one has
has
\be
\alpha_4^{KK} = \f {L_2^{KK}}{16 \pi^2}=
\f {\k^2 v^4}{8} \sum_{\vec n>0}\DD
=\f {v^4}{2M_S^4 (\delta -2)}
\ee
\bea
\alpha_5^{KK} &=&  \f {L_1^{KK}}{16 \pi^2}=
\f {\k^2 v^4}{8}\sum_{\vec n>0}\DD
\left (- \f 1 3 + \omega_\del^2 \f {\del -1}{2}\right )= -\f 1 {\del +2}
\f {\k^2 v^4}{8}\sum_{\vec n>0}\DD\nn\\
&=& -\f 1 {\del^2 -4}\f {v^4}{2M_S^4}
\label{a4a5kk}
\eea
Here we have also introduced the alternative parameterization \cite{gass}.

The dimension six remaining terms have been also considered
\cite{gene}. These 
contain anomalous fourlinear and
higher order gauge
couplings and $WW$ goldstone boson vertices. In the list of
possible operators of \cite{gene} these correspond
to $k_0^w$, $k_c^w$, $k_0^b$, $k_c^b$
terms and contribute to $WW\gamma\gamma$, $WWZ\gamma$ and
$ZZZ\gamma$ vertices 
\bea
&&\frac{k_0^w}{\Lambda^2}  g^2 \tr (\bW_{\mu \nu} \bW^{\mu \nu})
\tr (V^\alpha V_\alpha)
+\frac{k_c^w}{\Lambda^2} g^2 
\tr (\bW_{\mu \nu} \bW^{\mu \alpha}) \tr (V^\nu
V_\alpha)\nn\\
&&+\frac{k_0^b}{\Lambda^2}  g'^2 \tr (\bY_{\mu \nu} \bY^{\mu \nu})
\tr (V^\alpha V_\alpha)
+\frac{k_c^b}{\Lambda^2} g'^2 
\tr (\bY_{\mu \nu} \bY^{\mu \alpha}) \tr (V^\nu
V_\alpha)
\eea
with
\be
V_\alpha =(D_\alpha U)U^\dagger
\ee
Since
\be
\tr (V^\nu
V_\alpha) =-\tr [(D^\nu U)^\dagger (D_\alpha U)]
\ee
one has
\be
\frac{k_0^w}{\Lambda^2}  g^2= - \f {\kappa^2 v^2} 4
\sum_{\vec n>0}\DD = - \f {v^2}{4 M_S^4 (\del -2)}~~~~~
k_c^w=- 4  k_0^w
\ee
and 
\be
\frac{k_0^b}{\Lambda^2}  g'^2= - \f {\kappa^2 v^2} 4
\sum_{\vec n>0}\DD = - \f {v^2}{4 M_S^4 (\del -2)}~~~~~
k_c^b=- 4  k_0^b
\ee

In principle these  can be studied at LEP2 and future
linear colliders with
the processes
\begin{itemize}
\item $e^+e^-\rightarrow W^+W^-\gamma$
\item $e^+e^-\rightarrow Z\gamma\gamma$
\item $e^+e^-\rightarrow Z Z\gamma$
\end{itemize}
Present limits from LEP2 are  $k_{0,c}^{w,b}/\Lambda^2
\sim
10^{-2}-10^{-3}~GeV^{-2}$ 
\cite{gurtu} and increase to $k_{0,c}^{w,b}/\Lambda^2
\sim
10^{-5}~GeV^{-2}$ at a linear collider with
 $\sqrt{s}=500~GeV$ and an integrated luminosity of
$L=500~fb^{-1}$ \cite{gene}.
Similar bounds are obtained at LHC by studying the processes 
$pp\rightarrow \gamma\gamma W^*(\to l\nu)$ and 
$pp\rightarrow \gamma\gamma Z^*(\to ll)
$ \cite{ebo}.
However assuming
\be
 \kappa^2 \sum_{\vec n>0}\DD\sim \f 1{M_S^4}
\ee
and $M_S\sim 1~TeV$ the expected value of these anomalous couplings
is of order $k_{0,c}^{w,b}/\Lambda^2
\sim
10^{-7}-10^{-8}~GeV^{-2}$. This turns out much smaller than the possible
reaches of LHC and LC with $\sqrt{s}=500~GeV$.

The $\alpha_4$ and $\alpha_5$ terms contribute to $WW$ scattering
with strength which assuming  $M_S\sim 1~TeV$ are $O(10^{-3})$;
therefore they are at the border of the
regions which can be  be tested at LHC  with
$L=100~fb^{-1}$ and at a LC with
$\sqrt{s}=1.8~TeV$ and  $L=200~fb^{-1}$
\cite{zerw} (see also \cite{hky} for reviews on future
measurements  of electroweak symmetry breaking effective lagrangian
parameters). 

The dimension eight operators give additional contribution to
$WW$ scattering and furthermore contain a $\gamma^4$ coupling
which can be studied at $\gamma\gamma$ collider.
For instance at a $\gamma\gamma$ option of a 
$\sqrt{s}=500~GeV$ LC one can test $M_S\sim
4~TeV$ \cite{cheung}; at a $\sqrt{s}=1~TeV$ LC with
$L=200~fb^{-1}$
one can test masses $M_S\sim 3~TeV$
studying $e^+e^-\to \gamma\gamma 
e^+e^-$ \cite{atw}.

\end{document}